\documentclass[twocolumn]{revtex4}
\usepackage{graphicx}
\usepackage{dcolumn}
\usepackage{longtable}

\begin {document}

\title{Precise electromagnetic tests of ab-initio calculations of light nuclei: States in $^{10}$Be}

\author{E.A. McCutchan$^1$, C.J. Lister$^1$, R.B. Wiringa$^1$, Steven C. Pieper$^1$, D. Seweryniak$^1$, J.P. Greene$^1$, \\ M.P. Carpenter$^1$,  C.J. Chiara$^{1,2}$, R.V.F. Janssens$^1$, T.L. Khoo$^1$, T. Lauritsen$^1$, I. Stefanescu$^2$, S. Zhu$^1$}

\affiliation{$^1$ Physics Division, Argonne National Laboratory, Argonne, Illinois 60439, USA}

\affiliation{$^2$Department of Chemistry and Biochemistry, University of Maryland, College Park, Maryland 20742, USA}

\begin {abstract}
In order to test {\it ab-initio} calculations of light nuclei, we have remeasured lifetimes in $^{10}$Be using the Doppler Shift Attenuation Method (DSAM) following the $^{7}$Li($^7$Li,$\alpha$)$^{10}$Be reaction at 8 and 10 MeV. The new experiments significantly reduce systematic uncertainties in the DSAM technique. The J$^{\pi}$ = $2^+_1$ state at 3.37 MeV has $\tau$ = 205${\pm}$(5)$_{stat}\pm$(7)$_{sys}$ fs corresponding to a $B$($E$2$\downarrow$) of 9.2(3) $e^2$fm$^4$ in broad agreement with many calculations. The J$^{\pi}$ = $2^+_2$ state at 5.96 MeV was found to have a $B$($E$2$\downarrow$) of 0.11(2) $e^2$fm$^4$ and provides a more discriminating test of nuclear models. New Green's Function Monte Carlo (GFMC) calculations for these states and transitions with a number of Hamiltonians are also reported and compared to experiment. 

\end {abstract}

\maketitle

Several {\it ab-initio} approaches, {\it i.e.,} theoretical methods that start from ``bare''
nucleon-nucleon ($N\!N$) potentials (that reproduce 
elastic $N\!N$ scattering data) and empirical three-nucleon ($3N$) potentials,
are being developed for nuclear physics. Green's Function Monte Carlo (GFMC)~\cite{gfmc1,gfmc2} and the no-core shell model (NCSM)~\cite{ncsm1,ncsm2} are two methods that are used most to study 
p-shell nuclei. They have been successful in reproducing many features, 
including absolute binding energies and excitation spectra~\cite{pieper}, charge radii~\cite{mueller,Beradii}, and 
electromagnetic moments~\cite{pervin}. These methods are being developed as successors to the original nuclear shell model~\cite{inglis,kurath} and have provided new insight into the origin of the spin-orbit force, decoupling of poorly bound neutrons, 
and clustering of nucleons in light nuclei.  Several aspects of the 3$N$ force are not yet well constrained, particularly the isospin dependence.  This is important for the equation of state of neutron matter and the properties of neutron stars.  Thus, testing these approaches by studying some of the lightest nuclei can provide insights into some of the largest extended nuclear objects in the cosmos.

The $A$ = 10 nuclei $^{10}$Be, $^{10}$B, and $^{10}$C have long provided a stringent test of nuclear models. Kurath~\cite{kurath} discussed the need for an unusually strong spin-orbit interaction to correctly reproduce the sequence of levels in $^{10}$B. In these nuclei, the p-shell is half-filled and more than one state can be created with the same spin-parity $J^\pi$, isospin $T$, and spatial symmetry. For example, $^{10}$Be has two $J^\pi$=$2^{+}$, $T$=$1$ states at 3.37 and 5.96 MeV with dominant wave-function components $^1$D$_2$[442]. These states have very different properties and provide a delicate test of both the Hamiltonians and the calculational methods.  The Argonne $v_{18}$ (AV18) $N\!N$ \cite {wiringa} interaction predicts a separation of only 0.23(15) MeV for these states, one with a negative $(oblate)$ intrinsic quadrupole moment and a small $B$($E$2) decay to the ground state, and one with a positive $(prolate)$ quadrupole moment which is much more collective.  The oblate state is lower in energy with just AV18.  Calculations including 3$N$ potentials tend to separate the states and bring the more collective prolate-deformed level lower in energy. The ratio of electromagnetic decays from these states to the ground state is extremely sensitive to structure and varies from 4:1 to $>$50:1 in different nuclear models.  In this letter, we report on new experiments to precisely measure lifetimes of bound states in $^{10}$Be. New GFMC calculations with a number of realistic forces which explore the sensitivity to 3$N$ potentials are also presented.

To provide significant constraints on the new calculations implies achieving $<$5$\%$ accuracy in the $E$2 matrix elements, which is considerably higher than normally required for testing nuclear models.  The states of interest have short lifetimes, a few to hundreds of fs, so Coulomb excitation and the Doppler Shift Attenuation Method (DSAM) are the appropriate general experimental techniques. The study of $^{10}$Be following $\beta$-delayed neutron decay of $^{11}$Li has recently been shown to also be very powerful for this specific case \cite {fynbo,sarazin}.  We have concentrated on inferring lifetimes using DSAM, while focusing on the reduction of systematic uncertainties associated with the method. This technique relies on establishing the initial velocity of nuclei in the state of interest and then the mean velocity at decay after decelerating in a slowing material. From this velocity difference, the lifetime of the level can be determined if the slowing history of the ion is known.  Careful selection of the kinematic conditions for producing the states of interest at high velocities where stopping powers are well categorized, control of feeding from higher levels, and advancements in $\gamma$-ray detection all are important for moving beyond the original $>$25$\%$ measurements from the 1960s~\cite{warburton,fisher} and improving both precision and accuracy. 
  
Excited states in $^{10}$Be were populated in the $^7$Li($^7$Li, $\alpha$)$^{10}$Be reaction. Beams of $^7$Li ions of up to 50 pnA and energies of 8 and 10 MeV were produced by the ATLAS accelerator and impinged on thin $^7$Li metal and $^7$LiF targets. Target backings of copper and gold of sufficient thickness to slow the recoiling nuclei to about half their original energy were used. The nuclei recoiling along the beam direction were then detected in the Argonne Fragment Mass Analyzer (FMA) \cite{fma} positioned 90 cm downstream of the target and subtending $1^{\circ}\times 2^{\circ}$. The FMA rejected most non-interacting beam particles ($>10^{8}$ suppression) and transported  $^{10}$Be ions with charge state $q$=3$^+$ to the focal plane. The transmitted ions were identified 50 cm behind the focal plane in a 20 cm deep, two-electrode ion chamber operated at 15-30 torr. Gamma rays from the reaction were detected with the Gammasphere array~\cite{gammasphere}, consisting of 100 Compton-suppressed HPGe detectors in 16 azimuthally-symmetric rings mounted at $\theta$=$32^{\circ}$ to $163^{\circ}$ to the beam direction. The efficiency of the array (10$\%$ for 1-MeV $\gamma$ rays) and its granularity allowed many systematic tests to be made.

\begin{figure}
\center{{\includegraphics[height=120mm]{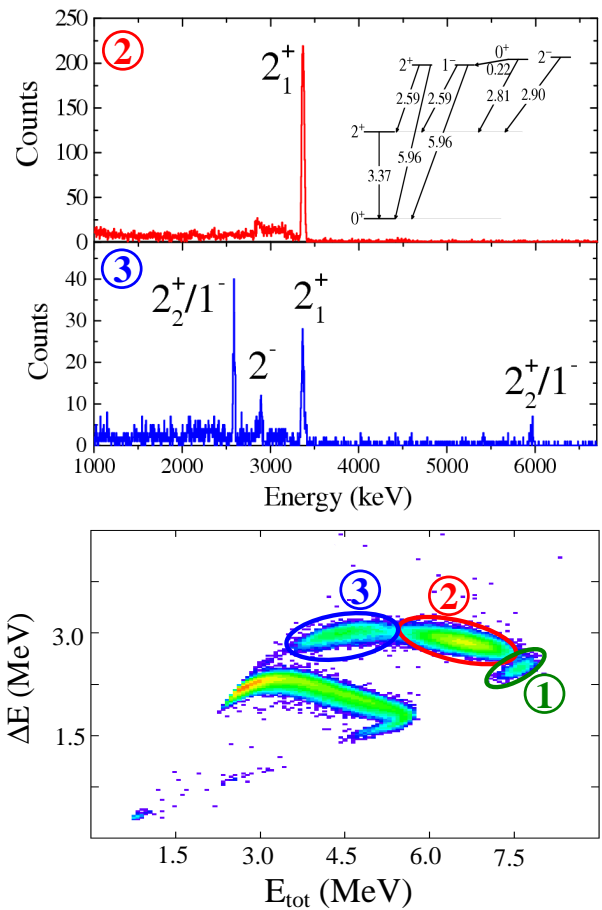}}}
\caption{(Color online) (Top and middle) Spectra gated on $^{10}$Be recoil groups with the $\gamma$-ray transitions labeled by the decaying state. (Bottom) Energy loss versus total energy from the ion chamber behind the focal plane of the FMA.  The circled regions correspond to direct population of distinct $^{10}$Be levels. The inset of the top panel shows a level scheme for the five bound excited states in $^{10}$Be with $\gamma$-ray energies in MeV.}
\end{figure}

The positive $Q$-value ($Q$=+11.4 MeV) gave $^{10}$Be nuclei high recoil velocities, and thus large Doppler shifts.
The velocity regime of $\beta$ = $v$/$c$ = 5-6$\%$ is ideal for DSAM measurements, as the stopping is 99.99$\%$ electronic and approaching the ``minimum-ionizing'' regime for nuclei stopping in materials. The current technique only sampled a small portion of the Bragg curves in the region where stopping is best known. The characterization of lithium and beryllium ions slowing in lithium, lithium fluoride, copper, and gold were taken from the SRIM package~\cite{srim}, which parameterizes the vast body of stopping data now available. Modern stopping powers are ~10$\%$ higher in this velocity regime than those used in the 1960s~\cite{northcliffe}.  The initial high recoil energy of $\sim$15 MeV is in contrast to the older experiments where it was $<$1 MeV and large-angle scattering influenced the direction of the emitting nucleus. 

The use of the FMA to select $^{10}$Be nuclei serves several purposes. The most obvious is to provide clean, background-free spectra.  In addition, due to the two-body kinematics of the reaction, a single recoil velocity vector is selected where only the $^{10}$Be nuclei traveling at 0$^{\circ}$ to the beam axis are studied. Finally, direct population of states can be selected, thereby eliminating cascade feeding.  In a plot of $\Delta E$ vs $E_{tot}$ for the ion chamber, given in Fig. 1, three distinct recoil groups are observed corresponding to direct population of the ground state (group 1), the $2_1^+$ state (group 2), and the set of four states around 6 MeV (group 3). The wide strip of counts below the $^{10}$Be recoils corresponds to $^7$Li scattered beam.  Gating on the direct population of the $2_1^+$ state (group 2), an exceptionally clean spectrum with a single $\gamma$ ray is observed (Fig. 1, top panel).  Selection of the lower energy recoils (group 3) gives the decay of the 6 MeV states, as presented in the middle panel of Fig. 1.    

For each state there is an optimum target layering, depending on the lifetime. For maximum sensitivity, equal numbers of decays should occur in the target production layer, in the slowing medium, and in the post-target flight region. Care was taken to monitor the thicknesses of targets and backings, using both weighing and $\alpha$-gauging, before and after the experiments, as this is a key source of systematic uncertainty. Several target materials and backings were used to investigate systematic effects.  Lithium metal targets with thickness of 100 - 200 $\mu$g/cm$^2$ were produced in an evaporator and transferred to the Gammasphere chamber in a vacuum interlock~\cite{lister} maintained at $<$10$^{-5}$ torr. These targets provided the simplest $\gamma$-ray spectra and allowed the use of high beam intensities, as there were few $\gamma$ rays produced in contaminant reactions. LiF targets were also used, to provide a contrasting stopping power in the target layer. However, reactions on the fluorine limited count rates, so these data sets were statistically inferior. 

The inference of lifetimes in the DSAM is obtained from determining the difference between production and emission velocities. If the deacceleration process is known, the mean lifetime of the state can then be deduced. The centroid of the 3.37-MeV, $2_1^+\rightarrow 0_1^+$ transition was measured in each of the 16 angle groups in Gammasphere.  The measured centroid as a function of cos($\theta$) for a copper target backing is given in Fig. 2(a), compared to a fit using the relativistic Doppler shift formula. Centroids were measured with sufficient precision that higher-order terms quadratic in cos($\theta$) were clear, and small mechanical mis-alignments could be corrected for.  The quality of the fit is more clearly illustrated in Fig. 2(b).  Here, the measured centroids are divided by the best fit function, $\sqrt{1-\beta^2}$ $/$ (1-$\beta$cos($\theta$)) with $\beta$=0.05109.   
Although the mean velocity of nuclei at the time of $\gamma$-ray emission produced in each target-backing combination varied widely, they all led to consistent lifetimes.  The mean lifetimes extracted from the analysis of the different backings are summarized in Table I. The calculation of the velocity-lifetime relationship was made by sub-dividing both the production target and backing into 30 layers and determining the average velocity profile and the probability of emission in each layer. The inferred lifetime did not change when a finer layering was used.  Both the statistical uncertainties (from the fit to the Doppler shifts) and the systematic uncertainties (from thickness and stopping power effects) varied considerably for the different target and backing arrangements. However, the many varied measurements allowed appraisal of systematic uncertainties.

\begin{figure}
\center{{\includegraphics[height=80mm]{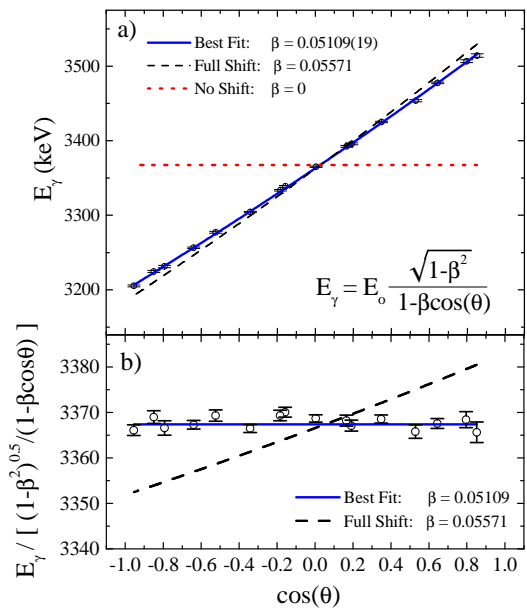}}}
\caption{(Color online) (a) Measured centroid of the 3.37-MeV transition for each of the Gammasphere angle groups compared to a fit with the relativistic Doppler formula.  Lines corresponding to maximum velocity of the $^{10}$Be recoils (dashed) and zero velocity (dotted) are shown. (b) Similar to (a) but normalized to the best fit function with $\beta$ = 0.05109(19).}
\end{figure}

\begin{table}
\caption{Mean lifetimes from different target and backing combinations
determined for the 3.37-MeV level in $^{10}$Be.}
\begin{tabular}{ccccccc}
\hline Target & $\;$ & Backing & $\;$ & $\tau$ & & $\;$ $\Delta$
$\tau_{stat}$ \\ ($\mu$g/cm$^2$) & & (mg/cm$^2$) & & (fs)& $\;$ & (fs)
\\ \hline 208 $\;$ $^7$Li & & 3.00 Cu & & 203 & & $\pm$9 \\ 208 $\;$
$^7$Li & & 2.66 Cu & & 209 & & $\pm$13 \\ 266 $\;$ $^7$LiF & & 2.28 Cu
& & 200 & & $\pm$23 \\ 208 $\;$ $^7$Li & & 2.07 Cu & & 189 & & +29
-26\\ 266 $\;$ $^7$LiF & & 2.16 Cu & & 204 & & +46 -37 \\ 90 $\;$
$^7$Li & & 2.69 Cu & & 208 & & $\pm$17 \\ 120 $\;$ $^7$Li & & 2.33 Au
& & 223 & & +67 -60 \\ \hline
\end{tabular}
\end{table}

The 3.37-MeV level was measured in seven separate experiments. The weighted mean value is $\tau$= 205${\pm}$(5)$_{stat}$ ${\pm}$(7)$_{syst}$ fs which gives a reduced transition probability $B$($E$2$\downarrow$)=9.2(3) $e^2$fm$^4$. Our lifetime is about 10$\%$ longer than the mean of previous experiments~\cite{tiley} which span the range 110 fs to 260 fs at the 1$\sigma$ level. The leading contribution to systematic uncertainty was the target layer thickness, especially for the lithium metal targets. In the future, Rutherford backscattering of the beam will be monitored during the experiment to determine the target thickness in live time at the point where the beam hits. The next leading term was the uncertainty in the stopping power of the slowing medium, which is now at the $<$ 2.5$\%$ level for this velocity regime \cite{srim}. 

Four bound states near 6 MeV were all populated and spectroscopic information was extracted. For this discussion, we focus on the second J$^{\pi}$=2$^{+}$ state at 5.96 MeV. This level is part of a close-lying doublet with a J$^{\pi}$=1$^{-}$ level that is only $\sim$1.5 keV higher and the $\gamma$ decays from the two states could not be resolved.  Thus, analysis of the 2.59-MeV and 5.96-MeV transitions yielded different ``effective'' lifetimes of 50(8) fs and 30(8) fs, respectively, highlighting the doublet nature of the transitions. 
Unfolding the doublets required knowledge of the $\gamma$-branching ratios from each level and their relative population intensities. Combining these two quantities and the measured ``effective'' lifetimes, the second 2$^{+}$ state was found to have $\tau$= 59${\pm}$(15) fs.  For the $2_2^+ \rightarrow 0_1^+$ branch, we used the precise value of 6.1(1)$\%$ measured in neutron capture~\cite{kennett}. In the decay to the ground state, this gives $B$($E$2$\downarrow$)=0.11(2) $e^2$fm$^4$.  Complete details of the analysis will be given in a future paper~\cite{longpaper}.

\begin{table*}
\caption{GFMC calculations of the $^{10}$Be ground-state energy $E_{gs}$, 
excitation energies $E_x$ (both in MeV), 
and  $B$($E$2$\downarrow$) transitions in $e^2$ fm$^4$.The calculations were done with a variety of potentials to explore the sensitivity in predicting electromagnetic matrix elements.}
\begin{ruledtabular}
\begin{tabular}{cddddd}
\multicolumn{1} {c}{$H$} &
\multicolumn{1} {c}{AV18} &
\multicolumn{1} {c}{AV18+UIX} &
\multicolumn{1} {c}{AV18+IL2} &
\multicolumn{1} {c}{AV18+IL7} &
\multicolumn{1} {c}{Expt.} \\
\hline
$|E_{gs}(0^+)|$ & 50.1(2) & 59.5(3)  & 66.4(4)  & 64.3(2) & 64.98 \\
$E_x(2^+_1)$  &  2.9(2) &  3.5(3)  &  5.0(4)  &  3.8(2) &  3.37 \\
$E_x(2^+_2)$  &  2.7(2) &  3.8(3)  &  5.8(4)  &  5.5(2) &  5.96 \\
$B(E2;2^+_1\rightarrow0^+)$
               & 10.5(3) & 17.9(5)  &  8.1(3)  &  8.8(2) &  9.2(3) \\
$B(E2;2^+_2\rightarrow0^+)$
                &  3.3(2) &  0.35(5)  &  3.3(2)  &  1.7(1) &  0.11(2) \\
$\Sigma B(E2)$	 & 13.8(4) & 18.2(6) & 11.4(4) & 10.5(3) & 9.3(3) \\

\end{tabular}
\end{ruledtabular}
\label{tab:qmc}
\end{table*}

In parallel to the experimental work, extensive GFMC calculations of states and transitions in $^{10}$Be were carried out using AV18 and a variety of 3$N$ forces, see Table II. The absolute GFMC energies are believed to be accurate to 1-2$\%$ for a given Hamiltonian, with somewhat larger errors for the excitation differences. The GFMC method for evaluating electromagnetic decays between states is detailed in Ref.~\cite{pervin}.  Calculations with the AV18 interaction alone predict near-degeneracy for the two J$^{\pi}$=2$^{+}$ states, with the oblate state slightly lower in energy.  To explore the sensitivity to 3$N$ potentials, calculations were performed with AV18 plus Urbana-IX (UIX), Illinois-2 (IL2), and Illinois-7 (IL7)~\cite{3body}.  Inclusion of the 3$N$ forces lifts the degeneracy between the $2^+$ states, slightly for UIX, more so for IL2 and approaching the experimentally observed spacing for the IL7 potential. 

Similar trends were found for the $B$($E$2) transition strengths to the ground state.  With just the AV18 potential, the calculations overpredict both 
transition strengths.  UIX provides a good description of the decay of the $2_2^+$ state, yet significantly overpredicts the decay of the $2_1^+$ state.  With the IL potentials, it appears that as the $2^+$ states become farther apart in energy, the predicted  $B$($E$2$\downarrow$) strengths tend to approach the experimental values.  Considering both the energies and transition strengths, the IL7 potential provides the best description.  However, while the summed $E$2 transition strength from the $2^+$ states to the ground state predicted by the IL7 potential is in very good agreement with experiment, the distribution of strength is not correct, with too much strength computed for the upper state. This shortcoming is still being investigated. It could possibly be remedied by further refinement of the Illinois three-body potentials, or by including more extended 2$^{+}$ components arising from particles in the sd-shell.

A wide variety of other models have also been used to investigate $^{10}$Be. In the traditional p-shell model with unmixed $K$-quantum number, Millener~\cite{millener} predicts $B$($E$2;$2^+_1 \rightarrow 0_1^+$)=11.8 $e^2$fm$^4$ and the ratio of $B$($E$2)s as 50:1. Both the NCSM with the CD-BONN, $NN$ potential~\cite{ncsm1} and the Microscopic Cluster Model (MCM)~\cite{cluster1} tend to underpredict the $B$($E$2; $2^+_1 \rightarrow 0_1^+$) decay, with NCSM predicting 6.6 $e^2$fm$^4$ and the MCM predicting 6.1 $e^2$fm$^4$.  However, both predict a weak $B$($E$2; $2^+_2 \rightarrow 0_1^+$) decay, $\sim$0.13 $e^2$fm$^4$, in good agreement with experiment. Thus, at present, no model provides a perfect description of $^{10}$Be and reconciling the predictions from these different models is an ongoing challenge.

In conclusion, the Doppler Shift Attenuation Method (DSAM) has been refined to precisely determine the lifetime of the first excited J$^{\pi}$=2$^{+}$ state in $^{10}$Be.  Care was taken to control the systematic uncertainties usually associated with DSAM.  High precision lifetime measurements now appear feasible in not only light nuclei, but also a range of systems which can be produced through two-body kinematic reactions.  The lifetime of the $2_2^+$ state in $^{10}$Be was also measured.  New {\it ab-initio} calculations show the sensitivity of the transition matrix elements to nuclear structure, especially to the form of three-body forces. In GFMC calculations the overall $B$($E$2) decay strength to the ground state can be well reproduced in calculations, but its distribution is still difficult to account for accurately.

This work was supported by the DOE Office of Nuclear Physics under contract DE-AC02-06CH11357, DE-FG02-94ER40834, and under SciDAC grant DE-FC02-07ER41457.  Calculations were performed on the SiCortex computer of the Argonne Mathematics and Computer Science Division.

\end {document}